\documentclass[11pt]{article}
\usepackage{amsmath}
\usepackage{amsfonts}
\usepackage{amssymb}
\usepackage{latexsym}
\usepackage{graphicx}
\usepackage{epsfig}
\usepackage{bm}
\usepackage[english]{babel}

\input epsf.sty

\begin{document}

{}~
{}~

\hfill
\vbox{
    \hbox{USTC-ICTS-07-19}
    }
\break

\vskip .4cm

\medskip

\baselineskip 20pt


\begin{center}

{\Large \bf Non-supersymmetric Attractors in Born-Infeld Black Holes
with a Cosmological Constant}

\end{center}

\vspace*{6.0ex}


\centerline{ \large Xian Gao}

\vspace*{4.0ex}

\centerline{ \it Interdisciplinary Center for Theoretical Study,
University of Science and Technology of China,}

\centerline{ \it Hefei, Anhui 230026, China}

\vspace*{1.0ex}

\centerline{\it and}

\vspace*{1.0ex}

\centerline{ \it Institute of Theoretical Physics, Academia Sinica,}

\centerline{ \it Beijing 100080, China}

\vspace*{1.0ex}

\centerline{ \tt gaoxian@itp.ac.cn}


\vspace*{5.0ex}


\begin{abstract}
We investigate the attractor mechanism for spherically symmetric
extremal black holes in Einstein-Born-Infeld-dilaton theory of
gravity in four-dimensions, in the presence of a cosmological
constant. We look for solutions analytic near the horizon by using
perturbation method. It is shown that the values of the scalar
fields at the horizon are only dependent on the charges carried by
the black hole and are irrelevant in their asymptotic values. This
analysis supports the validity of non-supersymmetric attractors in
the presence of higher derivative interactions in the gauge fields
part and in non-asymptotically flat spacetime.
\end{abstract}

\vfill \eject

\baselineskip 18pt



\section{Introduction}

The low energy limit of string theory gives rise to gravitational
model coupled to other fields, which typically have black hole
solutions. The black hole attractor mechanism has been an
interesting subject over the past several years, which states that
the near horizon geometry and field configurations turn out to be
completely independent of the asymptotic values of radially varying
moduli fields of the theory, especially, the moduli fields are
attracted to certain specific values at the horizon which are
dependent only on certain conserved quantities, such as charges
associated with the gauge fields and angular momentum. As a result,
the macroscopic entropy of the black hole is given only in terms of
these conserved charges and is independent of the asymptotic values
of the moduli.

The attractor mechanism was discovered first in $N=2$ BPS black
holes \cite{{9508072},{9602111},{9602136},{9603090}}. Later it was
found that the concept of attractor mechanism can also work in a
rather broad context, especially in non-supersymmetric cases
\cite{{9607108},{9702103}}. The non-supersymmetric attractors was
further clarified in \cite{0507096}. The authors considered theories
of gravity coupled to gauge fields and scalars in four and higher
dimensions which are asymptotically flat or AdS. Through
perturbative and numerical analysis of the full equations of motion,
it was shown that the attractor mechanism can work for
non-supersymmetric extremal black holes. The entropy function
formalism proposed by Sen \cite{0411255,0505122,0506177,0508042}, is
proved to be very useful in calculating the entropy of extremal
black holes in a general theory of gravity. By analyzing the
near-horizon field configurations, it is also shown that the
macroscopic entropy is independent of the asymptotic values of the
moduli, which implies the presence of attractor behavior.
Especially, it becomes clear that the attractor behavior is a
general phenomenon in extremal black holes, which have AdS as part
of their near-horizon geometries. Following these developments,
there has been a surge of interest in studying the attractor
mechanism without the use of supersymmetry, which shows that the
attractor mechanism can work for non-supersymmetric extremal black
holes, by analyzing the full solutions or by using entropy function
formalism
\cite{0511117,0512138,0510024,0601016,0602005,0606263,0606026,0608044,0606244,0611140,0611143,0704.1239,0706.1847}.
For a recent review of these developments, see \cite{0708.1270}.

The attractor mechanism has been proved to be very helpful to study
the properties of extremal black holes which are non-supersymmetric
solutions in supersymmetric theories and also solutions in theories
which have no supersymmetry. Especially, it can be very useful to
understand the structure of higher derivative terms in a general
theory of gravity
\cite{{0506177},{0508042},{0007195},{0009234},{0508218},{0602022},{0603149},{0604021},{0604106},{0608182}}.
It is well-known that the low energy limit of string theory gives
rise to the effective action of gravity which involves a variety of
higher derivative terms coming from both the gravity and the gauge
fields sides. The existence of non-supersymmetric attractor
mechanism in the presence of higher derivative terms has been
recently investigated in \cite{0602022}. A lot of interesting
aspects of Lovelock terms, Chern-Simons terms, Born-Infeld terms
etc., have been studied
\cite{0511306,0601228,0604028,0611240,0706.2046}.

The analysis of \cite{0507096} is based on the studying of the full
equations of motion of the metric, gauge fields and scalar fields
directly. However, in the absence of supersymmetry, the existence of
a full black hole solution interpolating between the near-horizon
geometry and the asymptotically infinity is non-trivial, especially
when there are higher derivative terms included. Furthermore, a full
low energy effective action of string theory has not been known yet.
Thus it is important to study non-supersymmetric attractor mechanism
when there are different kinds of higher derivative terms following
from the low energy limit of string theory, such as Gauss-Bonnet
term on the gravity side and Born-Infeld term on the gauge fields
side.

In recent years the Born-Infeld action has been occurring repeatedly
with the development of superstring theory, where the dynamics of
D-branes are governed by the DBI action. Extending the
Reissner-Nordstr\"{o}m black hole solutions in Einstein-Maxwell
theory to the charged black hole solutions in Einstein-Born-Infeld
theory with/without a cosmological constant has also attracted much
attention in recent years
\cite{wiltshire88,9702087,0004071,0004130,0007228,0101083,0301254,0408078,9903257,0306120,0406169,0410158}.
The attractor mechanism of black holes in Einstein-Born-Infeld
theory of gravity coupled to scalar fields has been studied in
\cite{0604028} by using entropy function formalism and in
\cite{0611240,0706.2046} by using effective potential for the
scalars and perturbation method. Using a perturbative approach to
study the corrections to the scalar fields and taking the
backreaction into the metric, it is shown that the scalar fields are
indeed drawn to fixed values at the horizon.

In this note, we generalize the result of
\cite{0604028,0611240,0706.2046} to the case in the presence of a
cosmological constant, where the spacetime is non-asymptotically
flat. Following the analysis in \cite{0604028,0611240,0706.2046}, we
show that the extremal EBI-AdS black hole solutions with regular
near-horizon configurations indeed exist and possess the attractor
behavior. In fact, most of the works concerning attractor mechanism
have been done in the asymptotically flat spacetime, therefore it is
very interesting to generalize the analysis to the
non-asymptotically flat case, especially asymptotic AdS. Thus due to
the AdS/CFT correspondence, one might be able to make a connection
between the attractor behavior of these AdS black holes and some
properties in the dual gauge theories.

This note is organized as follows. In section 2, we briefly review
the relevant features of attractor mechanism needed for our
purposes, following the outline of \cite{0507096}. In Section 3, we
study the Einstein-Born-Infeld theory of gravity coupled to scalar
fields, in the presence of a cosmological constant. In section 4, a
perturbative analysis is made to find possible extremal black hole
solutions. We discuss the existence of the solutions, calculate the
horizon radius and attractor values of the moduli fields. It is
shown that moduli fields indeed get attracted to fixed values at the
horizon. This result implies the presence of attractor mechanism in
Einstein-Born-Infeld-dilaton theory in the non-asymptotically flat
spacetime. Finally in section 5 we summarize our results.


\section{Brief Review of Non-supersymmetric Attractor Mechanism}

In this section we make a brief review of some relevant aspects of
non-supersymmetric attractors in four dimensional asymptotically
flat spacetime, following the analysis of \cite{0507096}. We
consider gravity coupled $U(1)$ gauge fields and scalars. The
scalars are coupled to gauge fields with dilatonic couplings. The
action has the form{\footnote{Here we choose the convention
$\epsilon_{\mu\nu\rho\sigma}=\sqrt{-G}\varepsilon_{\mu\nu\rho\sigma}$
and
$\epsilon^{\mu\nu\rho\sigma}=\frac{1}{\sqrt{-G}}\varepsilon^{\mu\nu\rho\sigma}$,
with $\varepsilon_{0123}=\varepsilon^{0123}=+1$.}}
\begin{equation}{\label{action_flat}}
    S=\frac{1}{\kappa^2}\int d^4 x \sqrt{-G}\left( R - 2\partial_{\mu}{\phi^i}\partial^{\mu}{\phi^i} -f_{ab}(\phi_i)F^a_{\mu \nu}F^{b\mu \nu} - \frac{1}{2}\tilde{f}_{ab}(\phi^i){\epsilon}^{\mu\nu\rho\sigma}F^a_{\mu \nu}F^b_{\rho
    \sigma}\right)\;,
\end{equation}
where $F^a_{\mu\nu},\;a=0,\cdots,N$ are $U(1)$ gauge fields and
$\phi^i,\;i=1,\cdots,n$ are scalar fields, $f_{ab}(\phi^i)$ and
$\tilde{f}_{ab}(\phi^i)$ determine the gauge couplings. It is
important that the scalars do not have a potential so that there is
a moduli space obtained by varying their values. However, we will
see that the coupling of these scalars with the gauge fields acts
like an ``effective potential'' for the scalars.

The equations of motion for the metric, dilatons and gauge fields
are derived from the action (\ref{action_flat}) as follows:
\begin{equation}
    R_{\mu\nu} - 2 \partial_{\mu}\phi^i\partial_{\nu}\phi^i =
    f_{ab}(\phi^i)\left( -2F^a_{\mu\lambda}F^{b\lambda}_{\phantom{ab}\nu} - \frac{1}{2}g_{\mu\nu} F^{a}_{\mu\nu}F^{b\mu\nu}\right)\;,
\end{equation}
\begin{equation}{\label{eq_dilaton_flat}}
    \frac{1}{\sqrt{-G}}\partial_{\mu}\left( \sqrt{-G} \partial^{\mu} \phi^i \right)
    = \frac{1}{4} \partial_i (f_{ab})
    F^{a}_{\mu\nu}F^{b\mu\nu} - \frac{1}{8} \partial_i ({\tilde{f}}_{ab}) \epsilon^{\mu\nu\rho\sigma}F^{a}_{\mu\nu}F^{b}_{\rho\sigma}\;,
\end{equation}
\begin{equation}
    \partial_{\mu} \left( \sqrt{-G} \left(f_{ab}(\phi^i)F^{b\mu\nu} + \frac{1}{2} \tilde{f}_{ab}(\phi^i)\epsilon^{\mu\nu\rho\sigma}F^{b}_{\rho\sigma} \right)\right) =
    0\;,
\end{equation}
in (\ref{eq_dilaton_flat}) $\partial_i \equiv {\partial}/{\partial
\phi^i}$. We also have the Bianchi identity for the gauge fields
\begin{equation}{\label{Bianchi}}
    \partial_{\mu}F^a_{\nu\rho} + \partial_{\nu}F^a_{\rho\mu} + \partial_{\rho}F^a_{\mu\nu} = 0\;.
\end{equation}

We consider static and spherically symmetric configurations. In
$3+1$ dimensions the metric and the gauge fields can be taken to be
of the form:
\begin{equation}{\label{metric_ansatz}}
    ds^2 = -{\alpha^2(r)} dt^2 + \frac{dr^2}{{\alpha^2(r)}} +
    {\beta^2(r)} d{\Omega}^2_2\;,
\end{equation}
\begin{equation}{\label{F_ansatz}}
    F^a = F^a_{tr}dt\wedge dr + F^a_{\theta\varphi}d\theta\wedge
    d\varphi\;.
\end{equation}
The equations of motion and the Bianchi identities for the gauge
fields can be solved directly by taking the gauge fields strengths
to be of the form:
\begin{equation}{\label{F_ansatz_new}}
    F^a =
    f^{ab}(\phi^i)(Q_{\mathrm{e}b}-\tilde{f}_{bc}(\phi^i)Q_{\mathrm{m}}^c)\frac{1}{\beta^2}dt\wedge
    dr + Q_{\mathrm{m}}^a\sin\theta d\theta \wedge d\varphi
    \;,
\end{equation}
where $Q_{\mathrm{e}}^a$ and $Q_{\mathrm{m}}^a$ are constants that
determine the electric and magnetic charges carried by the gauge
field $F^a$, and $f^{ab}$ is the inverse of $f_{ab}$. Under the
ansatz (\ref{metric_ansatz}) and (\ref{F_ansatz_new}), it is
possible to derive a set of second order differential equations for
$\alpha(r)$, $\beta(r)$ and $\phi(r)$ as follows:
\begin{align}\label{Einstein-Maxwell}
    \left(\alpha^2(r)\beta^2(r)£©\right)'' &=2\;,\\
    \left(\partial_r \phi^i\right)^2 + \frac{\beta''}{\beta}
    &=0\;,\\
    -1+\alpha^2 \beta'^2 + \frac{{\alpha^2}'{\beta^2}'}{2} &=
    -\frac{V_{\mathrm{eff}}(\phi^i)}{\beta^2} +
    \alpha^2\beta^2\left(\partial_r\phi^i\right)^2\;, {\label{constraint}}\\
    \partial_r\left( \alpha^2 \beta^2 \partial_r \phi^i \right) &=
    \frac{1}{2\beta^2}\frac{\partial V_{\mathrm{eff}}(\phi^i)}{\partial
    {\phi^i}}\;,
\end{align}
where (\ref{constraint}) is the first order Hamiltonian constraint
and the effective potential $V_{\mathrm{eff}}(\phi^i)$ is given by:
\begin{equation}{\label{Veff}}
    V_{\mathrm{eff}}(\phi^i) = f^{ab}(Q_{\mathrm{e}a} - \tilde{f}_{ac}Q_{\mathrm{m}}^c)(Q_{\mathrm{e}b} -
    \tilde{f}_{bd}Q_{\mathrm{m}}^d) + f_{ab}Q_{\mathrm{m}}^aQ_{\mathrm{m}}^b\;.
\end{equation}
The equations of motion given above can be derived from a
one-dimensional effective action:
\begin{equation}
    S= \frac{1}{\kappa^2} \int dr \left( 2-\left(\alpha^2\beta^2\right)'' -2\alpha^2\beta\beta'' - 2\alpha^2\beta^2(\partial_r\phi^i)^2 - \frac{2V_{\mathrm{eff}}(\phi^i)}{\beta^2}\right)\;,
\end{equation}
the Hamiltonian constraint (\ref{constraint}) must be imposed in
addition. We see that $V_{\mathrm{eff}}(\phi)$ plays the role of an
effective potential for the scalars.

We can now state two conditions which are sufficient for the
existence of an attractor \cite{0507096}. First, the charges should
be such that the resulting effective potential $V_{\mathrm{eff}}$,
as in (\ref{Veff}), has a critical point. We denote the critical
values for the scalars as $\phi^i(r)=\phi^i_0$, so that
\begin{equation}{\label{critical}}
    \left.\frac{\partial V_{\mathrm{eff}}(\phi^i)}{\partial
    \phi^i}\right|_{\phi^i=\phi^i_0} = 0\;.
\end{equation}
Second, the matrix of second derivatives of the effective potential
at the critical point,
\begin{equation}
    \left.M_{ij} \equiv \frac{\partial^2 V_{\mathrm{eff}}(\phi)}{\partial \phi^i \partial
    \phi^j}\right|_{\phi^i=\phi^i_0}\;,
\end{equation}
should have positive eigenvalues. Schematically we may write
\begin{equation}
    M_{ij}>0\;.
\end{equation}
This condition guarantees the stability of the solution. Once these
two conditions hold, it was argued in \cite{0507096} that the
attractor phenomenon results. Typically, there is a extremal black
hole solution in the theory, where the black hole carries the
charges determined by the paramters $Q_{\mathrm{e}}^a$ and
$Q_{\mathrm{m}}^a$. The moduli fields take critical values
$\phi^i_0$ at the horizon, which are independent of their values at
infinity, i.e., although $\phi^i$ are free at infinity as moduli
fields, they are attracted to fixed values $\phi^i_0$ at the
horizon.

As discussed in the introduction, the entropy function formalism
\cite{0411255,0505122,0506177,0508042} is a simple and powerful tool
to calculate the entropy of a extremal black hole in a general
theory of gravity, especially, the fact that the near-horizon field
configurations are determined by extremizing the entropy function
and the entropy is independent of the asymptotic values of the
scalars implies the presence of attractor mechanism. The entropy
function formalism focuses on the analysis of near-horizon
configurations, without known the full black hole solutions.
However, to see the moduli fields indeed get attracted to fixed
values when approaching the horizon, we have to use the formalism
for nonsupersymmetric attractor mechanism reviewed in this section,
which make explicit use of the general solutions and equations of
motion.


\section{Einstein-Born-Infeld-dilaton Theory with a Cosmological Constant}

We start with the following Einstein-Born-Infeld-dilaton action in
$3+1$ dimension in the presence of a cosmological constant
$\Lambda$:
\begin{equation}{\label{EBId+cc_Action}}
    S=\frac{1}{{\kappa}^2}\int d^4 x \sqrt{-G}\left[ R - 2\Lambda - 2 \partial_{\mu} \phi^i \partial^{\mu}
    \phi^i
    + {\mathcal{L}}_{BI}(F) \right]\;,
\end{equation}
where
\begin{equation}{\label{BI term}}
    {\mathcal{L}}_{BI}(F) = 4b^2 {s}^{-1} \left( 1-\sqrt{1+Y}
    \right),
\end{equation}
\begin{equation}{\label{def_Y}}
    Y \equiv \frac{s^2 F^2}{2b^2} - \frac{s^4}{16b^4}(F\ast F)^2\;.
\end{equation}
$b$ is the Born-Infeld parameter which has the dimension of mass,
$s=s(\phi^i)$ is the dilatonlike gauge coupling, and $F^2 \equiv
F_{\mu\nu}F^{\mu\nu}$, $F\ast F \equiv F_{\mu\nu} (\ast
F)^{\mu\nu}$. In (\ref{EBId+cc_Action}) for simplicity we consider
only one gauge field, if there are more than one gauge field, each
gauge field contributes a Born-Infeld term as (\ref{BI term}) but
with different dilatonlike gauge couplings. In string theory, the
Born-Infeld parameter $b$ is related to the string tension as
$b=\frac{1}{2\pi \alpha'}$. Note that when $b\rightarrow \infty$ the
Einstein-Born-Infeld theory reduces to the Einstein-Maxwell theory.

For simplicity, we restrict ourselves to the single scalar and gauge
field case. One can generalize the result to the case with several
scalars and gauge fields. We consider static and spherically
symmetric solution, thus we assume the metric and gauge field to be
of the form as (\ref{metric_ansatz}) and (\ref{F_ansatz}). We can
solve the gauge field first. Taking variation with respect to
$F_{\mu\nu}$ gives
\begin{equation}{\label{BI_F_eom}}
    \partial_{\mu}\left( \sqrt{-G}s^{-1} \frac{X^{\mu\nu}}{\sqrt{1+Y}} \right)=0\;,
\end{equation}
in which
\begin{equation}
    X^{\mu\nu} \equiv \frac{s^2 F^{\mu\nu}}{b^2} - \frac{s^4 (F\ast
    F){\epsilon}^{\mu\nu\rho\sigma}F_{\rho\sigma}}{8b^4}\;.
\end{equation}
We also have the Bianchi identity as (\ref{Bianchi}). Under the
static spherically symmetric ansatz, the equation of motion for the
gauge field (\ref{BI_F_eom}) and the Bianchi identity
(\ref{Bianchi}) can be solved as
\begin{equation}{\label{sol_F}}
    F_{tr} = \frac{s^{-1}Q_{\mathrm{e}}}{{\beta}^2\sqrt{1+\frac{Q_{\mathrm{e}}^2+Q_{\mathrm{m}}^2 s^2}{b^2
    {\beta}^4}}}\;, \qquad F_{\theta\varphi} = Q_{\mathrm{m}} \sin\theta\;.
\end{equation}
Here $Q_{\mathrm{e}}$ and $Q_{\mathrm{m}}$ are integration constants
and are related to the electric and magnetic charges carried by the
gauge field.

Taking variation of the metric gives the Einstein equation:
\begin{equation}{\label{Einstein_eqn}}
    R_{\mu\nu}-2\partial_{\mu}\phi \partial_{\nu}\phi -
    G_{\mu\nu}\Lambda = -2G_{\mu\nu}b^2 s^{-1}\left( 1-
    \frac{1}{\sqrt{1+Y}}\right) -
    \frac{2s}{\sqrt{1+Y}}F_{\mu\lambda}F^{\lambda}_{\phantom{a}\nu}\;,
\end{equation}
here $F_{\mu\nu}$ is given by (\ref{sol_F}). After substituting the
metric ansatz (\ref{metric_ansatz}), the
$R_{rr}-(G_{rr}/G_{tt})R_{tt}$ component of the Einstein equaton
(\ref{Einstein_eqn}) gives
\begin{equation}{\label{eom_1}}
    {\left(\partial_r \phi\right)}^2 + \frac{{\beta}''}{\beta}=0\;.
\end{equation}
The $R_{rr}$ component itself yields:
\begin{equation}{\label{eom_2}}
    {\left(\frac{{\alpha^2}' {\beta}^2}{2}\right)}' + {\beta}^2
    \Lambda = 2 b^2 \beta^2 s^{-1} \left( 1-\frac{1}{\sqrt{1+\frac{Q_{\mathrm{e}}^2+Q_{\mathrm{m}}^2 s^2}{b^2 \beta^4}}} \right)\;.
\end{equation}
Also the
$(R_{tt}-G_{tt}\Lambda)/(R_{\theta\theta}-G_{\theta\theta}\Lambda)$
component gives:
\begin{equation}{\label{eom_3}}
    -1 + {\left(\frac{\alpha^2 {\beta^2}'}{2}\right)}' +
    \beta^2\Lambda + \frac{1}{\beta^2}V_{\mathrm{eff}}(\phi)=0\;.
\end{equation}

Finally, the equation of motion for the scalar $\phi(r)$ takes the
form:
\begin{equation}{\label{eom_4}}
    \partial_r\left( \alpha^2 \beta^2 \partial_r \phi \right) =
    \frac{1}{2\beta^2}\frac{\partial V_{\mathrm{eff}}}{\partial
    {\phi}}\;,
\end{equation}
in which
\begin{equation}{\label{Veff_BI}}
    V_{\mathrm{eff}}(\phi) = 2b^2 \beta^4 s^{-1}\left( \sqrt{1+\frac{Q_{\mathrm{e}}^2+Q_{\mathrm{m}}^2 s^2}{b^2 \beta^4}} -1 \right)\;.
\end{equation}
We see that $V_{\mathrm{eff}}(\phi)$ plays the role of an
``effective potential'' for the scalar field. Here
$V_{\mathrm{eff}}(\phi)$, in contrast with Einstein-Maxwell theory,
is a function of $r$, as a result of to the nonlinearity of
Born-Infeld theory. However, as argued in \cite{0601016}, it is
possible to treat $r$ as just a parameter near the horizon.
Extremizing the effective potential and restricting the result to
the near-horizon region give the desired fixed values taken by the
moduli fields at the horizon.


\section{Perturbative Analysis}

In principle, one may suppose that if we indeed get a full black
hole solution of our theory with desired boundary conditions, we can
``see'' the attractor behavior directly from the $r$-dependance of
the dilaton fields $\phi(r)$. However, in a general theory of
gravity with gauge fields and scalar couplings, it is very difficult
to find a full set of exact solutions. On the other hand, to see the
attractor mechanism indeed exists, the near-horizon behavior of our
black hole solution is enough, even though the full solution is not
known.

A perturbative method was developed in \cite{0507096}. Generally,
the essential idea of the perturbative analysis is to start with an
extremal black hole solution of a the gravity, gauge fields and
scalars system, obtained by setting the asymptotic values of the
scalars equal to their critical values as in (\ref{critical}), then
examine what happens when the scalars take values at asymptotic
infinity which are somewhat different from their attractor values at
the horizon. From (\ref{Einstein-Maxwell})-(\ref{Veff}) we can see
that in common Einstein-Maxwell theory, if we set the asymptotic
values of the scalars to their critical values at the horizon, we
can set them to constants everywhere. Thus the equations of motion
(\ref{Einstein-Maxwell}) are much simplified, especially, we get a
extremal black hole solution, i.e., extremal
Reissner-Nordstr\"{o¡§}m black hole with constant scalars.

Thus one may suppose there is a similar extremal
Einstein-Born-Infeld black hole solution with constant-valued
moduli, which can be used as the starting point of the perturbative
analysis. Black hole solutions of Einstein-Born-Infeld theory
without any moduli fields have been constructed in
\cite{wiltshire88,9702087,0004071,0004130,0007228,0101083,0301254,0408078,9903257}
in asymptotic flat spacetime, and in \cite{0306120,0406169,0410158}
in the presence of a cosmological constant. In the absence of moduli
fields, the geometries are asymptotically flat and asymptotically
(A)dS respectively. However, in the presence of moduli fields, the
existence of a set of black hole solutions with desired boundary
conditions is highly non-trivial. From (\ref{eom_4}) and
(\ref{Veff_BI}) we can see that in contrast with
Einstein-Maxwell-dilaton theory, which holds a constant moduli as
its exact solution, due to the nonlinearity, the
Einstein-Born-Infeld-dilaton theory does not possess a black hole
solution with everywhere constant moduli{\footnote{It is well known
that the equations of motion admit $AdS_2 \times S^2$ as a solution
in the case of constant moduli.}}.

Thus, we take an analysis which is a little different from
\cite{0507096}. In view of the fact that the four equations of
motion (\ref{eom_1})-(\ref{eom_4}) are a set of highly complicated
coupled differential equations of order four, we follow the
Frobenius method to solve these equations as in
\cite{0604028,0611240,0706.2046}. We define $x \equiv
\left(\frac{r}{r_{\mathrm{H}}}-1\right)$ as the parameter of
expansion, i.e., we will find the solutions of $\alpha(r)$,
$\beta(r)$ and $\phi(r)$ in terms of $x$ order by order. We assume
the solutions to be extrmal, this guarantees the existance of the
attractor mechanism. Especially, we assume the solution has a
double-zero horizon{\footnote{As discussed in
\cite{0602005,0606263,0706.2046}, there is no attractor mechanism in
the single-zero horizon case. Some authors have shown that the
entropy function formalism also works well for some non-extemal
black holes, even though in general there is no attractor there
\cite{0701158,0703260,0705.2149}.}}, $\alpha^2(r) \equiv
(r-r_{\mathrm{H}})^2 {\tilde{\alpha}}^2(r)$ with
${\tilde{\alpha}}^2(r)$ being regular at the horizon
$r=r_{\mathrm{H}}$. Second, as a cosmological constant is included
in the theory, we assume the solution to be asymptotically (A)dS.
Also, we are interested in solutions which is regular at the
horizon, i.e. those with scalars do not blow up when approaching the
horizon.

We note that, from (\ref{Veff_BI}), $V_{\mathrm{eff}}(\phi)$ does
not have a minimum for any finite value of $\phi$ in the case of a
single electric or magnetic charge. In order to have a minimum in a
single charge case we need at least two gauge fields. On the other
hand, the non-existence of the extremal limit for electrically
charged black holes with Born-Infeld term was proposed in
\cite{0101083,0604028}. Thus we consider the dyonic case, with both
electric and magnetic charges are non-zero.

The most general Frobenius expansions of $\alpha(r)$, $\beta(r)$ and
$\phi(r)$ take the form as:
\begin{equation}{\label{fro_expansion}}
    \begin{aligned}
        {\alpha^2(r)} &=
        {\alpha}^2_{\mathrm{H}}x^2\sum^{\infty}_{m,n=0}a_{m,n}x^{m\lambda+n}\;,\\
        \beta(r) &= r_{\mathrm{H}}\sum^{\infty}_{m,n=0}b_{m,n}x^{m\lambda+n}\;,\\
        \phi(r) &=
        \sum^{\infty}_{m,n=0}{\phi}_{m,n}x^{m\lambda+n}\;,
    \end{aligned}
\end{equation}
In contrast with \cite{0507096,0611240} where $\lambda$ is assumed
to be $\lambda \ll 1$, here we assume $\lambda \geq 1$ in order to
guarantee $(\partial_r \phi)$ do not blow up at the horizon. From
the expansion(\ref{fro_expansion}), $\phi_0 = \phi(r_{\mathrm{H}})$
and so the moduli field is always fixed at the horizon, regardless
of any other information. Thus as argued in
\cite{0611240,0706.2046}, to complete the proof of the attractor
behavior, we should be able to show that the four sets of equations
of motion, denoting a coupled system of differential equations,
admit the expansions as (\ref{fro_expansion}). Furthermore, one
should see that there are solutions to all orders in the
$x$-expansion with arbitrary asymptotic values at infinity, while
the value at the horizon is fixed to be $\phi_0$. We should mention
that in the Einstein-Born-Infeld-dilaton theory, the existence of a
complete set of solutions with desired boundary conditions by itself
is not trivial.

Now let us take $s(\phi) = e^{-2\gamma \phi(r)}$, where $\gamma$ is
a parameter characterizing the coupling strength of dilaton field.
Note that in string theory $\gamma=1$.

\subsubsection*{Zeroth order results}

We start with a extremal black hole solution with double-zero
horizon at zeroth order perturbation:
\begin{equation}{\label{0th}}
    \phi_0(r)=\phi_0\;, \qquad \beta_0(r)=r_H\;, \qquad
    \alpha_0(r)={\alpha}_{\mathrm{H}}\left( \frac{r}{r_\mathrm{H}}-1
    \right)\;,
\end{equation}
we will see that $\phi_0$ is the attractor value of the dilaton,
$r_{\mathrm{H}}$ is the horizon radius. $\phi_0$, $r_{\mathrm{H}}$
and $\alpha_{\mathrm{H}}$ can be determined in terms of given
electric and magnetic charges. This can be done by substituting the
0-th order values of the fields (\ref{0th}) into the equations of
motion (\ref{eom_1})-(\ref{eom_4}). From the equation of motion for
the dilaton we get:
\begin{equation}{\label{phi0}}
    e^{-2\gamma\phi_0} = \frac{Q_{\mathrm{e}} \sqrt{Q_{\mathrm{e}}^2+b^2 {r_{\mathrm{H}}^4}}}{b
    Q_{\mathrm{m}}{r_{\mathrm{H}}}^2}\;.
\end{equation}
Note that we have the double horizon assumption, i.e.,
$\alpha^2(r_{\mathrm{H}})=0$ and ${\alpha^2}'(r_{\mathrm{H}})=0$,
then $r_{\mathrm{H}}$ can be solved from (\ref{eom_3}),
\begin{equation}{\label{r_H}}
    1 - \eta\frac{3}{\ell^2}{r_{\mathrm{H}}}^2 = \frac{1}{r^2_{\mathrm{H}}}V_{\mathrm{eff}}(\phi_0) = \frac{2bQ_{\mathrm{e}}Q_{\mathrm{m}}}{\sqrt{Q_{\mathrm{e}}^2+b^2
    {r_{\mathrm{H}}^4}}}\;,
\end{equation}
where we have parameterized $\Lambda = \eta\frac{3}{\ell^2}$, with
$\eta=-/+1$ for AdS/dS respectively. We can also solve
$\alpha_{\mathrm{H}}$ from (\ref{eom_2}):
\begin{equation}{\label{alpha_H}}
    \begin{aligned}
        \alpha_{\mathrm{H}}^2 &=\frac{2b^3{r^4_{\mathrm{H}}}Q_{\mathrm{e}}Q_{\mathrm{m}}}{(Q_{\mathrm{e}}^2+b^2{r^4_{\mathrm{H}}})^{3/2}}
        - \eta\frac{3}{\ell^2}{r_{\mathrm{H}}}^2\\
            &=
            \frac{b^2
            r^4_{\mathrm{H}}}{Q^2_{\mathrm{e}}+b^2r^4_{\mathrm{H}}}\left( 1-\eta\frac{3}{\ell^2}r^2_{\mathrm{H}} \right)-\eta\frac{3}{\ell^2}{r_{\mathrm{H}}}^2\;.
    \end{aligned}
\end{equation}
(\ref{phi0}) and (\ref{r_H}) together determine the attractor value
$\phi_0$ and horizon radius $r_{\mathrm{H}}$ in terms of the
charges, i.e., $\phi_0=\phi_0(Q_{\mathrm{e}},Q_{\mathrm{m}})$ and
$r_{\mathrm{H}}=r_{\mathrm{H}}(Q_{\mathrm{e}},Q_{\mathrm{m}})$.
Especially, due to (\ref{0th}) the Bekenstein-Hawking entropy is
determined by the electric and magnetic charges $Q_{\mathrm{e}}$ and
$Q_{\mathrm{m}}$.

We note that from (\ref{r_H}), the existence of a real positive root
of (\ref{r_H}), i.e., a extremal black hole with a double-zero
horizon of Einstein-Born-Infeld-dilaton theory in the presence of a
cosmological constant, is not always guaranteed. To analysis this
problem, we define $f(r_H)= 1 +
\eta\frac{3}{\ell^2}{r_{\mathrm{H}}}^2 -
\frac{2bQ_{\mathrm{e}}Q_{\mathrm{m}}}{\sqrt{Q_{\mathrm{e}}^2+b^2
{r_{\mathrm{H}}^4}}}$, thus the question becomes the existence of
positive roots of equation $f(r_{\mathrm{H}})=0$. In AdS case,
$\eta=-1$, we note that $f'(r_{\mathrm{H}}) =
\frac{6}{\ell^2}r^2_{\mathrm{H}}+\frac{4b^3Q_{\mathrm{e}}Q_{\mathrm{m}}r^3_{\mathrm{H}}}{(Q_{\mathrm{e}}^2+b^2
{r^4_{\mathrm{H}}})^{\frac{3}{2}}} > 0$, i.e., $f(r_{\mathrm{H}})$
is a monotonically increasing function of $r_{\mathrm{H}}$, and the
existence of positive $r_{\mathrm{H}}$ demands that
$f(0)=1-2bQ_{\mathrm{m}}<0$. Thus we find a lower bound for value of
the magnetic charge $2 bQ_{\mathrm{m}}>1 $ in
Einstein-Born-Infeld-dilaton theory in the presence of a negative
cosmological constant. Note that in AdS case with $\eta=-1$,
(\ref{alpha_H}) is always meaningful. Thus when this bound is
satisfied, (\ref{r_H}) has positive solution for $r_{\mathrm{H}}$,
i.e., a extremal black hole indeed exists{\footnote{Note that this
result is consistent with the proposal in \cite{0101083} about the
non-existence of the extremal limit for purely electrically charged
black holes in Einstein-Born-Infeld theories.}}. This bound indeed
relaxes in the limit $b \rightarrow \infty$. We focus on the case
with negative cosmological constant in the following discussion.


In the presence of a negative cosmological constant $\Lambda =
-{3}/{\ell^2}$, the exact expression of $r_H$ is complicated due to
(\ref{r_H}), which is a biquadratic algebraic equation for $r^2_H$.
In the limit $\ell \rightarrow \infty$, the result is simply
$r_{\mathrm{H}}={\left( 4Q^2_{\mathrm{e}}Q^2_{\mathrm{m}}
-Q^2_{\mathrm{e}}/b^2\right)}^{\frac{1}{4}}$, which is the horizon
radius in asymptotically flat spacetiem as given in
\cite{0611240,0706.2046}. In the limit $b \rightarrow \infty$,
Born-Infeld theory reduces to Maxwell theory, and $\phi_0$,
$\alpha_{\mathrm{H}}$ and $r_{\mathrm{H}}$ approach values in
Einstein-Maxwell-Dilaton theory in asymptotically AdS spacetime
\cite{0507096,0705.2892}. For example (\ref{r_H}) can be solved
perturbatively:
\begin{equation}
\begin{aligned}
    r^2_{\mathrm{H}} &= \frac{1}{6}\left(-\ell^2+\ell
    \sqrt{\ell^2+24Q_{\mathrm{e}}Q_{\mathrm{m}}}\right)\\
    &\qquad - \frac{1}{b^2}\frac{54Q^3_{\mathrm{e}}Q_{\mathrm{m}}}{\left(-\ell^2+\ell\sqrt{\ell^2+24Q_{\mathrm{e}}Q_{\mathrm{m}}}\right)\left(
\ell^2+30Q_{\mathrm{e}}Q_{\mathrm{m}}-\ell\sqrt{\ell^2+24Q_{\mathrm{e}}Q_{\mathrm{m}}}
\right)} + \cdots\;,
\end{aligned}
\end{equation}
the first term in the second line of the above expression is the
leading Born-Infeld correction to the horizon radius
$r^2_{\mathrm{H}}$ of the extremal Reissner-Nordstr\"{o}m-AdS black
hole in the large $b$ limit. Also, it is well-known that the
extremal RN-AdS black hole contains $AdS_2$ as part of its
near-horizon geometry. From (\ref{alpha_H}) we get
\begin{equation}{\label{alpha_H_corr}}
    \alpha^2_{\mathrm{H}} = 1+\frac{6}{\ell^2}r^2_{\mathrm{H}}
    -\frac{Q^2_{\mathrm{e}}}{b^2r^2_{\mathrm{H}}}\left( 1+\frac{3}{\ell^2}r^2_{\mathrm{H}} \right)
    +\cdots\;,
\end{equation}
again the third term in (\ref{alpha_H_corr}) can be understood as
the lowest order Born-Infeld corrections to $\alpha^2_{\mathrm{H}}$,
which is related to the size of $AdS_2$.

\subsubsection*{First order results}

At first order, we can write
\begin{equation}{\label{firstorder}}
\begin{aligned}
    \alpha^2(r) &=
    {\alpha}^2_{\mathrm{H}}x^2 +\delta{\alpha^2} \equiv {\alpha}^2_{\mathrm{H}}x^2(1+a_{1,0}x^{\lambda}+a_{0,1}x)\;,\\
    \beta(r) &= r_{\mathrm{H}} +\delta \beta \equiv r_{\mathrm{H}}(1+b_{1,0}x^{\lambda}+b_{0,1}x)\;,\\
    \phi(r) &= \phi_0+\delta \phi \equiv \phi_0 +
    \phi_{1,0}x^{\lambda}+\phi_{0,1}x\;.
\end{aligned}
\end{equation}
Substituting (\ref{firstorder}) into the equations of motion
(\ref{eom_1})-(\ref{eom_4}) and keeping $\delta(\alpha^2)$, $\delta
\beta$ and $\delta \phi$ as small parameters in perturbation, we get
linearized equations in terms of $\delta(\alpha^2)$, $\delta \beta$
and $\delta \phi$. Thus the undetermined coefficients $a_{1,0}$,
etc., and $\lambda$ can be read out from the expansions. From the
equation of motion for the scalar we get:
\begin{equation}{\label{phi_1st}}
    \delta \phi =
    \phi_{1,0}\left(\frac{r}{r_{\mathrm{H}}}-1\right)^{\lambda} +
    \phi_{0,1}\left(\frac{r}{r_{\mathrm{H}}}-1\right)\;,
\end{equation}
where $\phi_{1,0}$ is an undertermined constant, and
\begin{equation}{\label{phi_01}}
    \phi_{0,1} = \frac{\gamma\left(\alpha^2_{\mathrm{H}}-{3}r^2_{\mathrm{H}}/{\ell^2}\right)\left(1-\alpha^2_{\mathrm{H}}+{6}r^2_{\mathrm{H}}/{\ell^2}\right)}{\left(1+{3}r^2_{\mathrm{H}}/{\ell^2}\right)\left( (\gamma^2-1)\alpha^2_{\mathrm{H}} -\gamma^2 {3}r^2_{\mathrm{H}}/{\ell^2}
    \right)}\;,
\end{equation}
note that in the limit $\ell \rightarrow \infty$, the above result
indeed reduces to $\phi_{0,1} =
\gamma(1-\alpha^2_{\mathrm{H}})/(\gamma^2-1)$, which is the case in
asymptotically flat spacetime \cite{0611240,0706.2046}. $\lambda$
can also be determined as:
\begin{equation}{\label{lambda}}
    \lambda= \frac{1}{2}\left( -1+\sqrt{1+\frac{4 B^2}{
r^2_{\mathrm{H}}{\alpha^2_{\mathrm{H}}}}} \right)\;,
\end{equation}
with $B^2 \equiv \frac{1}{2} \frac{\partial^2
V_{\mathrm{eff}}}{\partial \phi^2}|_{\phi_0,r_{\mathrm{H}}}$.
Substituting (\ref{r_H}) and (\ref{alpha_H}) into (\ref{lambda}) we
get
\begin{equation}{\label{lambda_final}}
    \lambda = \frac{1}{2}\left( -1+\sqrt{1+8\gamma^2\left( 1- 3r^2_{\mathrm{H}}/{\ell^2} \right)}
    \right)\;,
\end{equation}
this result reduces to the asymptotically flat case as $\ell
\rightarrow \infty$, thus $\frac{B^2}{{\alpha^2_{\mathrm{H}}}
r^2_{\mathrm{H}}} = 2 \gamma^2$ and $\lambda = \frac{1}{2}\left(
 -1+\sqrt{1+8 \gamma^2}\right)$ as in \cite{0611240,0706.2046}.

As discussed in \cite{0611240,0706.2046}, in comparison to the
Einstein-Maxwell theory, here the metric get corrections at the
first order perturbation theory in $x$-expansion. Thus to this
order:
\begin{equation}
\begin{aligned}{\label{second_order}}
    \delta {\alpha}^2 &= \alpha^2_{\mathrm{H}} a_{1,0}
    \left( \frac{r}{r_{\mathrm{H}}} -1 \right)^{\lambda+2} + \alpha^2_{\mathrm{H}}
    a_{0,1} \left( \frac{r}{r_{\mathrm{H}}} -1 \right)^3\;,\\
    \delta \beta &= r_{\mathrm{H}} \left( \frac{r}{r_{\mathrm{H}}} -1
    \right)\;,
\end{aligned}
\end{equation}
where
\begin{equation}
\begin{aligned}{\label{a_2nd_corr}}
    a_{1,0} &= \frac{4\gamma\left( 1-\alpha^2_{\mathrm{H}}+6r^2_{\mathrm{H}}/{\ell^2} \right)\left(\alpha^2_{\mathrm{H}} - 3r^2_{\mathrm{H}}/{\ell^2}\right)}{(\lambda+1)(\lambda+2)\alpha^2_{\mathrm{H}}\left(1+3r^2_{\mathrm{H}}/{\ell^2}\right)} \phi_{1,0}\;,\\
    a_{0,1} &= \frac{2(\alpha^2_{\mathrm{H}}-3r^2_{\mathrm{H}}/{\ell^2})}{2{(1+3r^2_{\mathrm{H}}/\ell^2)}^2} \left\{ 1-\alpha^2_{\mathrm{H}} +\alpha^4_{\mathrm{H}} +\frac{3r^2_{\mathrm{H}}}{\ell^2}\left( 3+\alpha^2_{\mathrm{H}}+\frac{3r^2_{\mathrm{H}}}{\ell^2} \right) \right.\\
    & \qquad + \left.\gamma^2\left( 1-\alpha^2_{\mathrm{H}}+\frac{6r^2_{\mathrm{H}}}{\ell^2} \right)^2 \phi_{0,1}\right\} +\frac{2r^2_{\mathrm{H}}}{\ell^2} -\frac{4}{3}\;.
\end{aligned}
\end{equation}
Note that from (\ref{lambda_final}), the assumption $\lambda\geq1$
demands that $\gamma^2(1-3r^2_{\mathrm{H}}/\ell^2)\geq1$, this can
only be satisfied when $\gamma >1$. When this condition is
satisfied, this corrections (\ref{a_2nd_corr}) vanishes at the
horizon faster than $\left(r/r_{\mathrm{H}}-1\right)^2$, thus to
this order in $x$-expansion, the solution continues to be a double
horizon black hole with vanishing surface gravity.

\subsubsection*{Higher order results}

At the second order in $x$-expansion, the value of scalar field we
found at first order (\ref{phi_1st})-(\ref{phi_01}) plays the role
of a source, which results in corrections to the metric and the
scalar field itself. This can be calculated in a similar way as the
first order analysis.

In our perturbation analysis, we solve the equations of motion of
our system order by order in the $x$-expansion. As argued in
\cite{0602022,0611240,0706.2046}, we have seen that to the first
order, there is one parameter $\phi_{1,0}$ cannot be determined by
the equations of motion themselves. Let us denote the value of
$\phi_{1,0}$ as $K$. We thus find $a_{1,0}$ and $b_{1,0}$ as
functions of $K$. At any order $n\geq2$, we can substitute the
resulting values of $(a_{m,l},b_{m,l},\phi_{m,l})$, for all $m+l\leq
n$ from the previous orders. Thus
(\ref{eom_1}),(\ref{eom_3}),(\ref{eom_4}) of order $n$ and
(\ref{eom_2}) of order $(n-1)$ give
\begin{equation}
    a_{n,l}=a_{n,l}(K), \qquad b_{n,l}=b_{n,l}(K), \qquad
    \phi_{n,l}=\phi_{n,l}(K)\;,
\end{equation}
i.e., as polynomials of order $n$ in terms of $K$. $K$ remains a
free parameter to all orders in the $x$-expansions. From the
Frobenius expansion (\ref{fro_expansion}), all the coefficients are
functions of the single parameter $K$, thus the full black hole
solutions, especially the asymptotic values of $\alpha(r)$,
$\beta(r)$ and $\phi(r)$ are dependent on the parameter $K$. After
changing bases from $\left(\frac{r}{r_{\mathrm{H}}}-1\right)$ to
 $\left( 1-\frac{r_{\mathrm{H}}}{r}\right)$, it can be shown that $\alpha(\infty)$, $\beta(\infty)$ and $\phi(\infty)$ are free to
take different values given by different choices of $K$. The
convergence of this series should be addressed in detail, but it
should be convergent when $|K|$  is small enough. The fact that the
dilaton $\phi(r)$ can take arbitrary value at asymptotic infinity
$\phi(\infty)$ while its value at the horizon is fixed to be
$\phi_0$ as given in (\ref{phi0}) shows the presence of attractor
mechanism.


\section{Summary and Discussion}

In this note we studied attractor mechanism in Einstein-Born-Infeld
theory coupled to scalars and with dilaton-like gauge couplings, in
the presence of a cosmological constant in the action. We derived
the equations of motion of the system, and looked for possible
extremal black hole solutions with proper boundary conditions using
a perturbative method.

We focused on the case of asymptotic AdS black holes, which are more
interesting due to the AdS/CFT correspondence. We discussed the
existence of the extremal black hole solutions, calculated the
double-zero horizon radius and the attractor value of the dilaton.
It is shown that there are different extremal black hole solutions
characterized by different values of the scalars at asymptotic
infinity, while the scalar fields are indeed attracted to certain
fixed values at the horizon. This result generalizes the analysis in
\cite{0602022,0611240,0706.2046} and implies the presence of
attractor mechanism in the theory.

One can also study the case in asymptotic dS spacetime, though the
analysis of existence of such extremal black holes with desired
boundary condition is more complicated.


\bigskip


{\bf Acknowledgment:}

We would like to thank Miao Li, Da-Wei Pang, Tower Wang and
Zhao-Long Wang for useful discussions and kind help. We are grateful
to Miao Li for a careful reading of the manuscript and valuable
suggestions. This work was supported by grants from CNSF.


\end{document}